# The impact of the Hall term on tokamak plasmas

P.-A. Gourdain, Extreme State Physics Laboratory, Department of Physics and Astronomy, University of Rochester, NY 14627 USA

## Introduction

The Hall term has often been neglected in MHD codes as it is difficult to compute. Nevertheless, setting it aside for numerical reasons led to ignoring it altogether. This is especially problematic when dealing with tokamak physics as the Hall term cannot be neglected as this paper shows. In general, the Hall term is strongly expressed when the ion motion decouples from the electron motion [1]. This is often the case in tokamaks when non ambipolar flows are present [2]. This paper demonstrates the importance of the Hall effect by first deriving a global, dimensionless version of the generalized Ohm's law (GOL). While the standard dimensionless version of the GOL use local plasma quantities as scaling factors, this paper recast the GOL using global plasma quantities, such as average plasma density or vacuum toroidal field. In this approach, universal tokamak parameters like the Greenwald-Murakami-Hugill [3] density limit, the edge q factor [4] or the torus aspect ratio can be used to rescale the GOL, highlighting where the Hall term matters. After this short introduction, we derived in great detailed a dimensionless version of the GOL using global (i.e. average) plasma parameters. This scaling is done self-consistently with the other MHD equations, reducing the total number of free parameters. We then apply this dimensionless GOL to tokamak plasmas and show that the Hall term can rarely be neglected.

## The dimensionless generalized Ohm's law

The generalized Ohm's law is given by

$$\mathbf{E} = -\mathbf{u} \times \mathbf{B} + \eta \mathbf{J} + \frac{1}{en_e}[\mathbf{J} \times \mathbf{B} - \nabla.(p_e \mathbf{I})] + \frac{m_e}{n_e e^2}\left[\frac{\partial \mathbf{J}}{\partial t} + \nabla.\left(\mathbf{uJ} + \mathbf{Ju} - \frac{1}{en_e}\mathbf{JJ}\right)\right] \qquad (1)$$

The first term is the induction (or dynamo) electric field due to plasma flow perpendicular to the magnetic field direction. The second term is the Hall term electric field due to current flow perpendicular to the magnetic field direction.

It is usual to highlight the importance of each term of Eq. (1) by scaling this equation. We will use for this scaling, reference values which is the average values over the plasma volume (for E and B for instance) or characteristic values of plasma properties (such as L, the spatial scale length).

The most fundamental scaling is to drop space and time units using the reference scale length of the problem $L_0$ and the reference time $t_0$. As a result we get a reference velocity

$$u_0 = \frac{L_0}{t_0}. \tag{2}$$

It is more natural to use velocities rather than times as references. We will not use $t_0$ in the rest of this paper and substitute it with $L_0/u_0$. Starting from Maxwell's equations, we find that Faraday's law

$$\frac{\partial \mathbf{B}}{\partial t} = -\nabla \times \mathbf{E} \tag{3}$$

can be rescale using the reference quantities $B_0$, $L_0$, $t_0$ and $E_0$ as

$$\frac{u_0 B_0}{L_0} \frac{\partial \widetilde{\mathbf{B}}}{\partial \tilde{t}} = -\frac{E_0}{L_0} \widetilde{\nabla} \times \widetilde{\mathbf{E}}. \tag{4}$$

Now all quantities below the ~ symbol represent dimensionless quantities. Since $\widetilde{\mathbf{B}}$ and $\widetilde{\mathbf{E}}$ are the dimensionless magnetic and electric fields respectively they still satisfy the dimensionless Faraday's law

$$\frac{\partial \widetilde{\mathbf{B}}}{\partial \tilde{t}} = -\widetilde{\nabla} \times \widetilde{\mathbf{E}}. \tag{5}$$

Hence we get a relationship between the reference quantities $B_0$, $E_0$ and $u_0$

$$E_0 = B_0 u_0. \tag{6}$$

Using the same reasoning on Ampere's law

$$\nabla \times \mathbf{B} = \mu_0 \mathbf{J}, \tag{7}$$

we now define the reference current density

$$J_0 = \frac{B_0}{L_0 \mu_0}. \tag{8}$$

We need to also scale the other MHD equations so all reference quantities are consistent. The conservation of mass density equation

$$\frac{\partial \rho}{\partial t} + \nabla \cdot (\rho \mathbf{u}) = 0 \tag{9}$$

can be scaled using the using the reference ion mass density $\rho_0$ as

$$\frac{\rho_0 u_0}{L_0} \left[ \frac{\partial \tilde{\rho}}{\partial \tilde{t}} + \widetilde{\nabla} \cdot (\tilde{\rho} \widetilde{\mathbf{u}}) \right] = 0. \tag{10}$$

which does not give any new relationship between reference quantities. As $\rho_0$ does not appear in Eq. (1) and without any loss of generality, we suppose that the plasma has a single ion species of atomic mass $m_i$ so

$$n_0 = \frac{\rho_0}{m_i}. \tag{11}$$

Since Eq. (1) also supposes electro-neutrality, $n_0 = n_{i0} = n_{e0}$. In the rest of this paper we will use $n_0$ as a reference parameter instead of $\rho_0$. And since

$$\tilde{\rho} = \tilde{n}, \tag{12}$$

we will use $\tilde{n}$ instead. The conservation of momentum density equation

$$\frac{\partial \rho \mathbf{u}}{\partial t} + \nabla \cdot (\rho \mathbf{u}\mathbf{u} + p\mathbf{I}) = \mathbf{J} \times \mathbf{B} \tag{13}$$

can be rewritten as

$$\frac{\rho_0 u_0^2}{L_0} \left[ \frac{\partial \tilde{n}\tilde{\mathbf{u}}}{\partial \tilde{t}} + \widetilde{\nabla} \cdot (\tilde{n}\tilde{\mathbf{u}}\tilde{\mathbf{u}} + \tilde{p}\mathbf{I}) \right] = J_0 B_0 \tilde{\mathbf{J}} \times \tilde{\mathbf{B}} \tag{14}$$

This defines the reference pressure $p_0$ (i.e. reference energy density $e_0$) as

$$p_0 = \rho_0 u_0^2. \tag{15}$$

Finally, the conservation of energy density

$$\frac{\partial e}{\partial t} + \nabla \cdot \left( \left[ e + p + \frac{B^2}{2\mu_0} \right] \mathbf{u} \right) = \eta J^2 \tag{16}$$

becomes

$$\frac{p_0 u_0}{L_0} \left[ \frac{\partial \tilde{e}}{\partial \tilde{t}} + \widetilde{\nabla} \cdot \left( \left[ \tilde{e} + \tilde{p} + \frac{\tilde{B}^2}{2} \right] \tilde{\mathbf{u}} \right) \right] = \eta_0 J_0^2 \tilde{\eta} \tilde{J}^2, \tag{17}$$

giving the reference plasma resistivity

$$\eta_0 = \mu_0 L_0 u_0 \tag{18}$$

And the reference magnetic field $B_0$

$$\frac{B_0^2}{\mu_0} = p_0. \tag{19}$$

Using Eqs. (15) and (19), the plasma reference velocity $u_0$ has to be

$$u_0 = \frac{B_0}{\sqrt{m_i n_0 \mu_0}}. \tag{20}$$

which is the reference Alfvén speed of the plasma. The local dimensionless plasma Alfvén speed,

$$\tilde{u}_A = \frac{u_A}{u_0} = \frac{\tilde{B}}{\sqrt{\tilde{n}}}, \tag{21}$$

and the local plasma sound speed,

$$\tilde{c}_s = \frac{c_s}{u_0} = \sqrt{\frac{\gamma \tilde{p}}{\tilde{n}}}, \tag{22}$$

can be used to rescale the plasma velocity using the local Alfvén Mach number $M_A$

$$M_A = \frac{u}{u_A} = \frac{\tilde{u}}{\tilde{u}_A}, \qquad (23)$$

or the thermal Mach number $M_{th}$

$$M_{th} = \frac{u}{c_s} = \frac{\tilde{u}}{\tilde{c}_s}. \qquad (24)$$

As usual the plasma beta,

$$\beta = \frac{2\mu_0 p}{B^2} = \frac{2\tilde{p}}{\tilde{B}^2}, \qquad (25)$$

connects the plasma Alfvén speed to the sound speed using the relation

$$\frac{c_s}{u_A} = \frac{\tilde{c}_s}{\tilde{u}_A} = \sqrt{\frac{\gamma}{2}\beta}. \qquad (26)$$

Using Eqs. (22) and (24) we can rewrite Eq. (14) as

$$\frac{\partial \tilde{n}\tilde{u}}{\partial \tilde{t}} + \tilde{\nabla}.[\tilde{p}(\gamma M_{th}^2 \mathbf{vv} + \mathbf{I})] = \tilde{\mathbf{J}} \times \tilde{\mathbf{B}} \qquad (27)$$

where $\mathbf{v}$ is the unit velocity vector,

$$\mathbf{v} = \frac{\tilde{\mathbf{u}}}{\|\tilde{\mathbf{u}}\|}. \qquad (28)$$

In the end, we are left with only three free reference parameters, namely $L_0$, $B_0$ and $n_0$.

We can rewrite Eq.(1) using all dimensionless quantities, indicated by the ~,

$$\tilde{\mathbf{E}} = -\tilde{\mathbf{u}} \times \tilde{\mathbf{B}} + A_1 \tilde{\eta}\tilde{\mathbf{J}} + \frac{A_2}{\tilde{n}_e}\tilde{\mathbf{J}} \times \tilde{\mathbf{B}} - \frac{A_3}{\tilde{n}_e}\tilde{\nabla}.(\tilde{p}_e \mathbf{I}) - \frac{A_4}{\tilde{n}_e}\tilde{\nabla}.\left(\frac{1}{\tilde{n}_e}\tilde{\mathbf{J}}\tilde{\mathbf{J}}\right) + \frac{A_5}{\tilde{n}_e}\left[\frac{\partial \tilde{\mathbf{J}}}{\partial \tilde{t}} + \tilde{\nabla}.\left(\tilde{\mathbf{u}}\tilde{\mathbf{J}} + \tilde{\mathbf{J}}\tilde{\mathbf{u}}\right)\right]. \qquad (29)$$

According to the previous equations the values of the scaling parameters are

$$A_1 = \frac{\eta_0}{L_0 u_0 \mu_0}, A_2 = \frac{B_0}{en_0 L_0 u_0 \mu_0}, A_3 = \frac{p_0}{en_0 u_0 L_0 B_0}, A_4 = \frac{m_e B_0}{e^3 n_0^2 u_0 L_0^3 \mu_0^2}, A_5 = \frac{m_e}{e^2 n_0 L_0^2 \mu_0} \qquad (30)$$

From Eq. (18), $A_1$ is 1 and Eq. (19) $A_3 = A_2$. Using the electron inertial length $\delta_e = (m_e/\mu_0 e^2 n_e)^{1/2}$, the length at which the electrons decouple from the magnetic field, gives

$$A_4 = A_2 \frac{\delta_{e0}^2}{L_0^2}, A_5 = \frac{\delta_{e0}^2}{L_0^2}. \qquad (31)$$

Using Eq. (20) and the ion inertial length, $\delta_i = (m_i/\mu_0 e^2 Z^2 n_i)^{1/2}$ we get

$$A_2 = \frac{\delta_{i0}}{ZL_0}. \qquad (32)$$

The ion inertial length is the length at which the ion motion decouples from the electron motion and the electron remain frozen in the magnetic field1. The rescale resistivity

$$\tilde{\eta} = \frac{\eta}{\eta_0} = \frac{\eta}{L_0 u_0 \mu_0} = \frac{1}{\tilde{S}} \qquad (33)$$

is simply the inverse Lundquist number. Thus, the generalized Ohm's law can be rescaled as:

$$\widetilde{\mathbf{E}} = -\tilde{\mathbf{u}} \times \tilde{\mathbf{B}} + \frac{1}{\tilde{S}} \tilde{\mathbf{J}} + \frac{\delta_{i0}}{L_0} \frac{1}{\tilde{n}_i} \left[ \tilde{\mathbf{J}} \times \tilde{\mathbf{B}} - \widetilde{\nabla} \cdot (\tilde{p}_e \mathbf{I}) \right] + \frac{\delta_{e0}^2}{L_0^2} \frac{1}{\tilde{n}_e} \left[ \frac{\partial \tilde{\mathbf{J}}}{\partial \tilde{t}} + \widetilde{\nabla} \cdot \left( \tilde{\mathbf{u}} \tilde{\mathbf{J}} + \tilde{\mathbf{J}} \tilde{\mathbf{u}} - \frac{\delta_{i0}}{L_0} \frac{1}{\tilde{n}_e} \tilde{\mathbf{J}} \tilde{\mathbf{J}} \right) \right]. \qquad (34)$$

where all reference plasma parameters are consistent with MHD and Maxwell's equations. All reference parameters are using the subscript 0. They correspond to plasma global, average parameters. Local dimensionless plasma properties use the ~ symbol, except for usual dimensionless numbers such as Mach numbers or plasma beta. Due to the absolute ratio of the electron to the ion mass, the electron inertial terms can be dropped and we finally get

$$\widetilde{\mathbf{E}} = -\tilde{\mathbf{u}} \times \tilde{\mathbf{B}} + \frac{1}{\tilde{S}} \tilde{\mathbf{J}} + \frac{\delta_{i0}}{L_0} \frac{1}{\tilde{n}_i} \left[ \tilde{\mathbf{J}} \times \tilde{\mathbf{B}} - \widetilde{\nabla} \cdot (\tilde{p}_e \mathbf{I}) \right] + \mathcal{O}\left( \frac{\delta_{i0}}{\tilde{n}_e L_0} \right). \qquad (35)$$

## Generalized Ohm's law ordering in tokamaks

In tokamak plasmas our reference (i.e. average) parameters are $L_0 = a$, where a is the plasma minor radius, $B_0 = B_T$, where $B_T$ is the toroidal magnetic field at the plasma major radius R. The ratio R/a is the plasma aspect ratio A. The reference plasma density $n_0$ is taken to be the Greenwald-Murakami-Hugill [3,5] line average density limit

$$n_0 = 10^{20} \frac{I_P(MA)}{\pi L_0^2} \qquad (36)$$

For a circular tokamak with edge safety factor $q_a$ and aspect ratio A, the plasma current is given by [4]

$$I_P(MA) = \frac{5 L_0 B_0}{A q_a}. \qquad (37)$$

Thus, the reference density becomes

$$n_0 = 10^{20} \frac{5 B_0}{\pi A q_a L_0}. \qquad (38)$$

Now that all free reference parameters $n_0$, $L_0$ and $B_0$ are defined it is possible to get the different scaling parameters in Eq.(35). We first look at the scaling of the dynamo term for tokamaks. Plasma flows have low Alfvénic Mach numbers (on the order of $10^{-2}$ for spontaneous plasma rotation [6]), $\tilde{\mathbf{u}}$ is typically small. Using Eqs. (21) and (26) we can rewrite $\tilde{\mathbf{u}}$ using the local Alfvén Mach number

$$\tilde{\mathbf{u}} = \frac{|u|}{u_0} \mathbf{v} = \frac{|u|}{u_A} \frac{u_A}{u_0} \mathbf{v} = M_A \frac{\tilde{B}}{\sqrt{\tilde{n}}} \mathbf{v}. \qquad (39)$$

Then we focus on the resistive scaling term $\tilde{\eta}$. We choose the Spitzer resistivity model to carry out our equation ordering:

$$\tilde{\eta} = 2.4 \times 10^{-5} \frac{\ln\Lambda}{T_e^{3/2}(\text{eV})} \frac{1}{\sqrt{Aq_a L_0^3(\text{m}) B_0(\text{T})}} \tag{40}$$

Finally, we focus on the scaling term $A_2$ for Hall and battery effects. First we compute the reference Alfvén speed using Eq.

$$u_0 = 10^{-10} \sqrt{\frac{\pi A q_a L_0 B_0}{5 m_i \mu_0}}. \tag{41}$$

After defining the reference Alfvén speed, the scaling parameter of the Hall term in Eq.(35) can be computed

$$A_2 = 10^{-10} \sqrt{\frac{\pi A q_a m_i}{5 e^2 L_0 B_0 \mu_0}}. \tag{42}$$

Once again, we can simplify its value to

$$A_2 \simeq 1.8 \times 10^{-2} \sqrt{\frac{Aq_a}{L_0(\text{m}) B_0(\text{T})}}. \tag{43}$$

In low-beta tokamaks, the electrical current density is mostly aligned with the toroidal field. Thus, we can scale the value of the dimensionless current density $\|\tilde{\mathbf{j}}\|$ by using its toroidal component. We use the large aspect ratio approximation of the safety factor q and magnetic shear s [7]

$$q \simeq \frac{2\tilde{B}}{\langle \tilde{j} \rangle A} \text{ and } s = 2\left(1 - \frac{\tilde{j}}{\langle \tilde{j} \rangle}\right), \tag{44}$$

where $\langle \tilde{j} \rangle$ is the average current density enclosed by the flux surface numbered by the value q, to get

$$\tilde{\mathbf{j}} \simeq \frac{2\tilde{B}}{Aq}\left(1 - \frac{s}{2}\right)\mathbf{j}, \tag{45}$$

where

$$\mathbf{j} = \frac{\tilde{\mathbf{j}}}{\|\tilde{\mathbf{j}}\|}. \tag{46}$$

From Eqs. (39) and (45), the dimensionless generalized Ohm's law for tokamaks becomes

$$\widetilde{\mathbf{E}} \simeq -M_A \frac{\tilde{B}}{\sqrt{\tilde{n}}} \mathbf{v} \times \tilde{\mathbf{B}} + 1.8 \times 10^{-2} \frac{1}{\tilde{n}} \sqrt{\frac{Aq_a}{L_0(\text{m}) B_0(\text{T})}} \left[ \frac{2\tilde{B}}{Aq}\left(1 - \frac{s}{2}\right) \mathbf{j} \times \tilde{\mathbf{B}} - \widetilde{\nabla}.(\tilde{p}_e \mathbf{I}) \right] \\ + 2.4 \times 10^{-5} \frac{1}{\sqrt{Aq_a L_0^3(\text{m}) B_0(\text{T})}} \frac{\ln\Lambda}{T_e^{3/2}(\text{eV})} \tilde{\mathbf{j}}. \tag{47}$$

The electric field ordering previously discussed demonstrates that the resistive term can be dropped in modern tokamaks, where the temperature is above 10 eV. Further, the nature of the Hall effect does not affect resistive electric fields, which are aligned with $\tilde{\mathbf{j}}$. Since electrical current and plasma flows are on flux surfaces to order $r_L/L_0$, where $r_L$ is the ion gyroradius [8], the non-resistive part of the electric field is largely perpendicular to the purely resistive part of the electric field. Thus, we will ignore the resistive electric field in the rest of this paper.

To evaluate the impact of the Hall term on the plasma electric field, we take the ratio HD of the Hall scaling factor to the dynamo scaling factor:

$$\text{HD} \simeq 3.6 \times 10^{-2} \sqrt{\frac{q_a}{AL_0(m)B_0(T)}} \frac{(1-\frac{s}{2})}{qM_A\sqrt{\tilde{n}}}. \tag{48}$$

This factor has contributions from global plasma properties (first square root in Eq. (48)), which in modern tokamaks ($L_0 \sim 1$, $B_0 \sim 3$, $A \sim 2$, $q_a \sim 3$) is on the order of 0.7. Further since $M_A$ is below 0.04 in high performance regimes [9], the HD factor is bounded as $\text{HD} > \text{HD}_{\text{min}}$ where:

$$\text{HD}_{\text{min}} = \frac{2(1-\frac{s}{2})}{3q\sqrt{\tilde{n}}}. \tag{49}$$

In the plasma core ($q \sim 1$, $s \sim 0$, $\tilde{n} \sim 2$) $\text{HD}_{\text{min}}$ is on the order of 0.5. For equilibria with reverse shear ($s<0$) this ration is even larger. Near the plasma edge ($q \sim 3$, $s \sim 1.5$, $\tilde{n} \sim 0.1$) $\text{HD}_{\text{min}}$ is above 0.2. Thus, the Hall term cannot be neglected compared to dynamo in the electric field generated in tokamaks.

Modern views on tokamak equilibria have demonstrated the importance of inertial terms on plasma dynamics [10,11]. We just argued that the Hall term cannot be ignored and should be included in equilibrium computations as well. We believe it is an important contributor in the formation of the pressure pedestal in H-modes. In this high confinement mode of the plasma, the negative radial (i.e. pointing inward, in the **JxB** direction) electric field is accompanied by the formation of a steep pressure profile [12,13,14]. In the H-mode pedestal $n_e \sim n_i$ and $T_e \sim T_i$, $\nabla p_e \sim \nabla p_i$ [15]. Further basic equilibrium considerations dictate that $\nabla p_i$ points inwards. Thus, the electric field generated by the $\nabla p_e$ term in the GOL points outward and cannot account for the inward (i.e. negative) electric field observed in the H-modes. Only the Hall effect and dynamo are left to consider. For a plasma where the inertial term has been neglected, Eq. (13) becomes

$$\nabla p = \mathbf{J} \times \mathbf{B}. \tag{50}$$

The well-known equilibrium equation can be used in Eq. (1), where the electron inertia and resistive terms have also been neglected,

$$\mathbf{E} = \mathbf{u} \times \mathbf{B} - \frac{1}{en}\nabla p_i. \tag{51}$$

We supposed electro-neutrality ($n_i = Zn_e$) and the fact that $p = p_i + p_e$. Eq. (51) corresponds exactly to Eq. (5) from Hinton and Staebler [16], where inertial terms have also been neglected. In their framework, the H-mode electric field connects only to the ion momentum balance.

## Conclusion

This paper showed that the dimensionless generalized GOL can be rendered dimensionless using global parameters instead of local ones as it usually the case. We also demonstrated that when this rescaling is done is conjunction with other MHD equations, the reference velocity used must be the plasma Alvén speed. Further by including tokamak scaling into the GOL, we demonstrated that the Hall term cannot be neglected in tokamaks. The large radial electric fields found in H-modes can be fully explained by the inclusion of the Hall in the GOL. In conclusion, the Hall term cannot be ignored if one wants to properly capture plasma dynamics in tokamak MHD modeling.